\documentclass[aps,preprint,prd]{revtex4}
\usepackage{graphics}
\usepackage{latexsym}


\begin{document}

\title{The neutrino star in the bulk Universe}
\author{Ryszard Ma\'{n}ka}
\email{manka@us.edu.pl}
\homepage{www.cto.us.edu.pl/~manka}
\author{D.Karczewska}
\email{dkarcz@us.edu.pl} 
\affiliation{University of Silesia, Institute of Physics, Katowice 40007,
ul. Uniwersytecka 4, Poland}
\date{\today}

\begin{abstract}
Motivated by the Kaluza-Klein theory with large extra spacetime dimensions the
neutrino star built from the massive sterile neutrinos core and the massless
brane neutrinos envelope is presented. The six-dimensional compactification
scale \( \sim 15 \) MeV gives maximal neutrino mass \( M_{max}=2.3\times 10^{4}\, M_{\odot } \)with
radius \( R_{max}=1.2\times 10^{6}\, km \). The maximal neutrino star parameters
varies with temperature. In the limit of the neutrino ball approximation the
maximal sterile neutrino star is \( M_{max}=1.1\times 10^{6}\, M_{\odot } \).
\end{abstract}
\maketitle
 
\section*{Introduction}
Recently there has been considerable interest in the field theories with large
extra spacetime dimensions. In the comparison to the standard Kaluza-Klein theory
these extra dimensions may be restricted only to the gravity sector of theory
while the Standard Model (SM) fields are assumed to be localized on the 4-dimensional
spacetime \cite{1},\cite{4},\cite{ADD3}.This is a promising scenario from
the phenomenological point of view because it shift the energy scale of unification
from \( 10^{19}\, GeV \) to \( 1-100\, TeV \). It has been recently shown
\cite{L1} that this framework can be embedded into string models, where the
fundamental Planck scale can be identified with the string scale which could
be as low as the weak scale. The extra dimensions have the potential to lower
the unification scale as well \cite{DDG1}.

The aim of this paper is to examine the degenerate neutrino star originated
from the extra dimensional theory. The neutrino star ( neutrino ball ) was a
subject of interest in the theory of the Standard Model \cite{hd}. The fermion
star in the extra dimensional theory was also the a subject of interest \cite{nk}.The
neutrino star model is capable to explain the nature of the object in Sgr A\( ^{*} \)
in the center of the Galaxy \cite{torr}.

\section*{The bulk neutrino extension of the electroweak theory}
Much of the interesting phenomenology of brane-world models is associated with
the Kaluza-Klein theories \cite{tap} that originate from large, gravity-only
additional dimensions. The higher-dimensional bulk fermions lead to Kaluza-Klein
towers of standard model singlets that may be interpreted as sterile neutrinos
\cite{dinest,ADDM,dvali,BCS}. In this paper we shall consider the six-dimensional
Kaluza - Klein theory. Let us now consider the action in the six-dimensional
spacetime: 
\begin{equation}
\label{a1}
{\mathcal{S}}=\int d^{6}x\sqrt{-g_{6}}L={\mathcal{S}}_{g}+{\mathcal{S}}_{F}=\int d^{6}x\sqrt{-g_{6}}(L_{g}+L_{F}),
\end{equation}
 where \( g_{6}=det(g_{MN}) \) and \( M=\{\mu ,\, i\} \), \( N=\{\nu ,\, j\} \)
with \( x^{M}=\{x^{\mu },y^{i}\},\, i=1,2 \). The metrical tensor in the six-dimensional
spacetime can be written: 
\begin{equation}
\label{a2}
g_{MN}=\left( \begin{array}{cc}
e^{-2\xi (x)/f_{0}}\overline{g}_{\mu \nu } & 0\\
0 & -\delta _{ij}e^{+2\xi (x)/f_{0}}
\end{array}\right) .
\end{equation}
 According to the above definition we can write: 
\begin{equation}
\label{a3}
\sqrt{-g_{6}}=\sqrt{-\overline{g}}e^{-2\xi (x)/f_{0}}.
\end{equation}
 We consider here the Lagrangian of the field as follows:
\begin{eqnarray}
 & L=L_{g}+L_{F}, & \\
 & L_{g}=-\frac{1}{2\kappa _{6}}R, & \\
 & L_{F}=L_{brane}\delta (y)+L_{bulk}, & 
\end{eqnarray}
 where \( \kappa _{6} \) is the six-dimensional gravitational coupling. Its
natural interpretation originates from the distance scaling in the four-dimensional
spacetime. Let us compactify the six-dimensional spacetime \( ({\mathcal{M}}_{6}\rightarrow {\mathcal{M}}_{4}\times {\mathcal{S}}^{1}\times {\mathcal{S}}^{1}) \)
to the four-dimensional Minkowski one. In this paper we assume that the extra
dimensions are compactified on the \( 2 \) dimensional torus with a single radius
\( r_{2} \). The six-dimensional action may be rewritten as: 
\begin{equation}
\label{a7}
{\mathcal{S}}=\int d^{4}x\int d^{2}y\sqrt{-g_{6}}L=\int d^{4}x\sqrt{-\overline{g}}{\mathcal{L}},
\end{equation}
 where \( \int d^{2}y=(2\pi r_{2})^{2} \). The six-dimensional gravitional
coupling \( \kappa _{6}=8\pi G_{6} \) is convenient to define as 
\[
G_{6}^{-1}=\frac{4\pi }{(2\pi )^{2}}M^{4},\]
 where \( M \) - is the energy scale of the compactification (\( \sim 10-100\, TeV \)).
Cosmological consideration \cite{hall} gives the bound \( M>100\, TeV \)
what corresponds \( r_{2}<5.1\times 10^{-5}\, mm \). If we define the four-dimensional
coupling constant \( \kappa =8\pi G_{N}=8\pi M^{-2}_{Pl} \) we get 
\begin{equation}
\label{mpl2}
M^{2}_{Pl}=4\pi M^{4}r_{2}^{2}.
\end{equation}
The parameter \( f_{0} \) is defined as: 
\begin{equation}
f^{2}_{0}=\frac{1}{\pi }M^{2}_{Pl}
\end{equation}
to produce the \( 1/2 \) term in for the dilaton field kinetic term. The model
can be motivated by the ten-dimensional string theory with the string scale \( M_{s} \).
For simplicity we assume that the first \( 2 \) dimensions have the same size
characterized by the radius \( r_{2} \). This is likewise assumed for the remaining
\( 4 \) dimensions where the corresponding radius is called \( r_{4}<r_{2} \).
The four-dimensional Planck constant
\begin{equation}
(\frac{M_{Pl}}{M_{s}})^{2}=16\pi (2\pi r_{2}M_{s})^{2}(2\pi r_{4}M_{s})^{4}
\end{equation}
is connected to the string scale \( M_{s} \). In the result of the comparison to
(\ref{mpl2}) we have 
\begin{equation}
M=\sqrt{4\pi }(2\pi r_{4}M_{s})M_{s}.
\end{equation}
The paremeters of the model \cite{karmen} are presented on the Table \ref{tab1}.
\begin{table}
{\centering \begin{tabular}{|c|c|c|c|c|}
\hline 
&
\( M_{s} \)&
\( M_{2}=r_{2}^{-1} \)&
\( r_{4}^{-1} \)&
\( M \)\\
\hline 
\hline 
&
\( 1.2 \) TeV&
\( 1.6 \) MeV&
\( 432.05 \) MeV&
\( 7.42\, 10^{7} \) GeV\\
\hline 
&
\( 1.2 \) TeV&
\( 15 \) MeV&
\( 141.11 \) MeV&
\( 2.27\, 10^{8} \) GeV\\
\hline 
\end{tabular}\par}

\caption{\label{tab1}The paremeters set of the model \cite{karmen}.}
\end{table}
In this section we shall extend the Standard Model minimally with the bulk neutrino
\( \psi (x,y) \). The lagrangian of the neutrino sector of the model is then:
\[
L_{brane}=i\overline{L}_{f}\widehat{\Gamma }^{\mu }\partial _{\mu }L_{f}-\frac{1}{M_{s}}(h_{f,a}\overline{\psi }_{a}L_{f}H_{0}+h.c.),\]
 and the fermion field are
\begin{equation}
L_{f}=\left( \begin{array}{c}
\nu ^{0}_{f}\\
0
\end{array}\right) _{L}\, \, \, f=e,\nu ,\tau .
\end{equation}
The Higgs field will have only the residual form 
\begin{equation}
H_{0}=\frac{1}{\sqrt{2}}\left( \begin{array}{c}
0\\
\upsilon 
\end{array}\right) 
\end{equation}
after the spontaneous symmetry breaking. The additional bulk neutrino is described
by the Lagrange function
\begin{eqnarray}
\mathcal{L}_{bulk} & = & i\overline{\psi }\widehat{\Gamma }^{N}\partial _{N}\psi -m_{D}(\overline{\psi }_{L}\psi _{R}+h.c),
\end{eqnarray}
 where ~\( \psi _{R,L}=(I\pm \Gamma ^{7})\psi  \). The system has global \( U(1) \)
lepton symmetry which generates the lepton charge \( Q_{L} \). The bulk neutrino
field can be decomposed into four-dimensional Dirac spinors \( \psi _{a} \),
where \( a=\{+,-\} \). The gamma matrices \( \Gamma  \) are defined as 
\begin{equation}
\{\widehat{\Gamma }^{M},\widehat{\Gamma }^{N}\}=2g^{M,N}I.
\end{equation}
In the flat four-dimensional spacetime, when \( \overline{g}_{\mu \nu }=\eta _{\mu \nu } \)
they may be defined as 
\begin{equation}
\label{ga1}
\widehat{\Gamma }^{\mu }=e^{\xi /f_{0}}\Gamma ^{\mu },\, \, \, \Gamma ^{\mu }=\gamma ^{\mu }\otimes I_{(2)},
\end{equation}
\[
\widehat{\Gamma }^{i}=e^{2\xi /f_{0}}\Gamma ^{i},\, \, \, \Gamma ^{i}=i\gamma ^{5}\otimes \sigma ^{i}\]
and
\begin{eqnarray*}
\{\gamma ^{\mu },\gamma ^{\nu }\}=2\eta ^{\mu \nu }I, &  & \\
\{\gamma ^{i},\gamma ^{j}\}=-2\delta _{i,j}I, &  & 
\end{eqnarray*}
where 
\[
\Gamma ^{7}=\overline{\Gamma }^{0}...\overline{\Gamma }^{5}=\gamma ^{5}\otimes \sigma ^{3},\, \, \, \, \gamma ^{5}=-i\gamma ^{0}\gamma ^{1}\gamma ^{2}\gamma ^{3}.\]
Using the metric tensor six-dimensional form (\ref{a2}) and the (\ref{ga1})
we can calculate the Lagrange function in the four-dimensional Minkowski spacetime
\begin{eqnarray}
L=i\overline{L}\gamma ^{\mu }D_{\mu }L+e^{-2\xi /f_{0}}\sum _{a,\underline{n}}i\overline{\psi }_{a,\underline{n}}(x)\gamma ^{\mu }D_{\mu }\psi _{a,\underline{n}}(x) &  & \\
-m_{D}\sum _{a,\underline{n}}\overline{\psi }_{a,\underline{n},L}\psi _{a,\underline{n},R}-e^{\xi /f_{0}}\sum _{a,\underline{n}}\overline{\psi }_{a,\underline{n}}(x)(\gamma ^{i}u_{\underline{n}}^{i})\psi _{a,\underline{n}}(x), &  & \nonumber 
\end{eqnarray}
where
\begin{equation}
u_{\underline{n}}^{i}=\frac{1}{r_{2}}n^{i},\, \, \, \, m^{2}_{\underline{n}}=\underline{u}^{2}_{\underline{n}}=m\frac{1}{r_{2}^{2}}(n_{1}^{2}+n_{2}^{2}).
\end{equation}
The bulk neutrino is decomposed as
\begin{equation}
\psi _{a}(x,y)=\frac{e^{\xi /f_{0}}}{(2\pi r_{2})}\sum _{\underline{n}}\psi _{a,\underline{n}}(x)exp(\frac{1}{r_{2}}i\underline{n}\underline{y}).
\end{equation}
 Each of these four-dimensional Dirac spinors can be decomposed into Weyl spinors
\begin{equation}
\psi _{a,\underline{n}}(x)=\left( \begin{array}{c}
\xi _{a,\underline{n}}^{c}\\
\eta _{a,\underline{n}}
\end{array}\right). 
\end{equation}
After electroweak symmetry breaking we introduce the mass 
\begin{equation}
m_{f,a}=\frac{h_{f,a}v}{2\pi (M_{s}r_{2})}\sim 10^{-1}\, \, MeV.
\end{equation}
 The exact diagonalization of the mass matrix gives three exactly massles Weyl
fermions
\begin{equation}
\nu _{f}=U_{f,f'}\nu ^{0}_{f'}+V_{f,a}(\underline{0})\eta _{a}(\underline{0})
\end{equation}
and two Dirac spinors for each mode number \( \underline{n} \) with mass 
\begin{equation}
M_{\underline{n}}=\sqrt{m_{D}^{2}+m^{2}_{\underline{n}}}.
\end{equation}
In general \cite{karmen}, there is a superposition of the electroweak neutrinos
\( \nu ^{0}_{f} \) on the brane and the Kaluza-Klein bulk neutrinos \( \eta _{a}(\underline{n}) \).

\section*{The neutrino star}
An alternative model for the supermassive compact object in the center of our
Galaxy has been recently proposed by Tsiklauri and Viollier \cite{rm}\cite{tsi}.
The main ingredient of the proposal is that the dark matter at the center of
the galaxy is non-baryonic composed with massive neutrinos or gravitinos. Such
neutrino balls could have formed in early epochs, during a first-order phase
transition in the Standard Model. 

In case of the spherically symmertic gravitational field we have:
\begin{equation}
\label{a5}
\overline{g}_{\mu \nu }=\left( \begin{array}{cccc}
e^{\nu (r)} &  &  & \\
 & -e^{\lambda (x)} &  & \\
 &  & -r^{2} & \\
 &  &  & -r^{2}\sin ^{2}\theta 
\end{array}\right) .
\end{equation}
In the similar way the dilaton filed \( \xi (r) \) will be dependent on the
radius \( r \). As \( \xi (r)\sim f_{0}=M_{Pl}/\sqrt{\pi }\approx 10^{19}\, GeV \)
we shall neglect the dilaton filed \( \xi (r) \) in the first approximation.
In this paper we present numerical results describing the structure of neutrino
star. It is possible to describe a static spherical star solving the
Oppenheimer-Tolman-Volkoff (OTV)
equations (more general case with the dilaton filed \( \xi (r) \) is presented
in Appendix I). 
\begin{equation}
\label{teq1}
\frac{dP(r)}{dr}=-\frac{G}{r^{2}}(\rho
(r)+\frac{P(r)}{c^{2}})\frac{(m(r)+\frac{4\pi }{c^{2}}P(r)r^{3})}{(1-\frac{2Gm(r)}{c^{2}r})},
\end{equation}
\begin{equation}
\label{teq2}
\frac{dm(r)}{dr}=4\pi r^{2}\rho (r).
\end{equation}
 Having solved the OTV equation the pressure \( p(r) \), mass \( m(r) \) and
density \( \rho (r) \) were obtained. To obtain the total radius \( R \) of
the star the fulfillment of the condition \( p(R)=0 \) is necessary. Our aim
is to achieve the equation of state of neutrino star matter at finite temperature.
In such a case the physical system can be defined by the thermodynamic potential
\( \Omega  \) \cite{fet}
\begin{equation}
\Omega =-k_{B}TlnTr(e^{-\beta (H-\mu Q_{L})}),
\end{equation}
 where \textbf{\( \beta =1/k_{B}T \)}, \( k_{B} \) is the Boltzmann constant,
\( Q_{L} \) lepton charge. The chemical potential \( \mu  \) reflects the
lepton number conservation. \( H \) stands for the Hamiltonian of the physical
system. All needed averages are calculated with the Hamiltonian \( H \). We
define the density of energy and pressure by the energy - momentum tensor 
\begin{equation}
T_{\mu \nu }=(P+\epsilon )u_{\mu }u_{\nu }-Pg_{\mu \nu },
\end{equation}
 where \( u_{\mu } \) is a unite vector (\( u_{\mu }u^{\mu }=1 \)). So, the
calculations give 
\begin{equation}
\epsilon (x_{F},T)=\rho c^{2}=\epsilon _{F,bulk}+\epsilon _{F,brane},
\end{equation}
\begin{equation}
P(x_{F},T)=P_{F,bulk}+P_{F,brane},
\end{equation}
 where 
\begin{equation}
\epsilon _{F}=\epsilon _{0}\chi (x_{F},T),
\end{equation}
\begin{equation}
P_{F}=P_{0}\phi (x_{F},T).
\end{equation}
 The fact that the bulk Dirac neutrinos are massive means that they play the
same role like ions in a white dwarf or neutrons in a neutron star. We have 
\begin{eqnarray}
\chi _{bulk}(x_{F},T)=\frac{1}{^{\pi ^{2}}}\int _{0}^{\infty }dz\, z^{2}\sqrt{z^{2}+1}\{\frac{1}{\exp ((\sqrt{1+z^{2}}-\mu ')/\tau )+1} &  & \\
+\frac{1}{\exp ((\sqrt{1+z^{2}}+\mu ')/\tau )+1}, &  & \nonumber 
\end{eqnarray}
\begin{eqnarray}
\phi _{bulk}(x_{F},T)=\frac{1}{3\pi ^{2}}\int _{0}^{\infty }\frac{z^{4}dz}{\sqrt{z^{2}+1}}\{\frac{1}{\exp ((\sqrt{1+z^{2}}-\mu ')/\tau )+1} &  & \\
+\frac{1}{\exp ((\sqrt{1+z^{2}}+\mu ')/\tau )+1}\}, &  & \nonumber 
\end{eqnarray}
 where \( \tau =(k_{B}T)/M_{2} \), 
\begin{equation}
\label{mu}
\mu '=\mu /M_{2}=\sqrt{1+x_{F}^{2}}
\end{equation}
 and 
\begin{equation}
\label{mui}
x_{F}=k_{F}/M_{s}.
\end{equation}
 Similarly to the paper \cite{toki} we have introduced (\ref{mu},\ref{mui}) the
dimensionless ``Fermi'' momentum even at finite temperature which exactly
corresponds to the Fermi momentum at zero temperature. For the massless brane
neutrinos we have 
\begin{eqnarray}
\chi _{brane}(x_{F},T)=\frac{\gamma }{2\pi ^{2}}\int dz\, z^{3}\{\frac{1}{(e^{(z-x_{F})/\tau }+1)}+\frac{1}{(e^{(z+x_{F})/\tau }+1)}\} &  & \\
=-\frac{3\gamma }{\pi ^{2}}t^{4}\int ^{\infty }_{0}\frac{dz\, z^{3}}{e^{(z-x_{F})/\tau }+1}\left\{ Li_{4}(e^{x_{F}/\tau })+Li_{4}(e^{-x_{F}/\tau })\right\} , &  & 
\end{eqnarray}
 where the presure is
\begin{equation}
\phi _{brane}(x_{F},T)=\frac{1}{3}\epsilon _{brane}(x_{F},T).
\end{equation}
 Both \( \epsilon _{F} \) and \( P_{F} \) depend on the neutrino chemical
potential \( \mu  \) or Fermi momentum \( x_{F} \). This parametric dependence
on \( \mu  \) (or \( x_{F} \)) defines the equation of state. 

Similarly  to the paper \cite{toki} we have introduced the dimensionless 'Fermi' momentum
even at finite temperature which exactly corresponds to the Fermi momentum at
zero temperature. Both \( \epsilon _{F} \) and \( P_{F} \) depend on the neutrino
chemical potential \( \mu  \) or Fermi momentum \( x_{F} \). This parametric
dependence on \( \mu  \) (or \( x_{F} \)) defines the equation of state. 

\begin{figure}
{\par\centering \resizebox*{10cm}{!}{\includegraphics{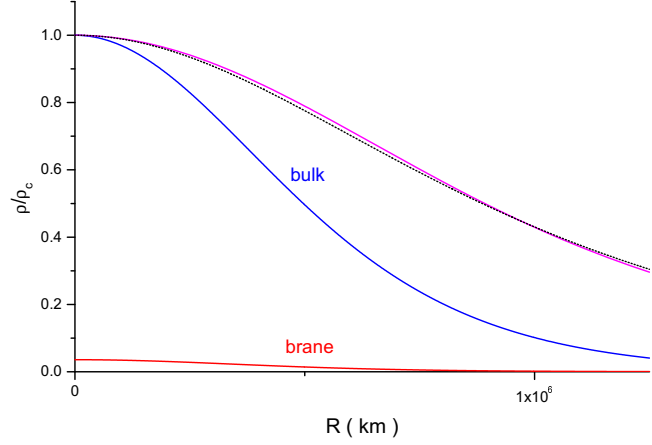}} \par}
\caption{\label{rfig1}The density profile of the neutrino star for the bulk neutrino
(blue), brane ones (red), in the ultrarelativistic limit (magenta) and in the
neutrino ball case (the dot line).}
\end{figure}

\begin{figure}
{\par\centering \resizebox*{10cm}{!}{\includegraphics{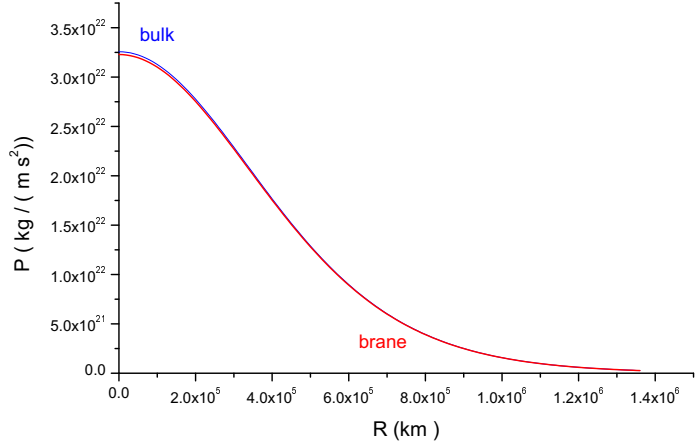}} \par}
\caption{\label{rfig2}The pressure profile of the neutrino star.}
\end{figure}

\begin{figure}
{\par\centering \resizebox*{10cm}{!}{\includegraphics{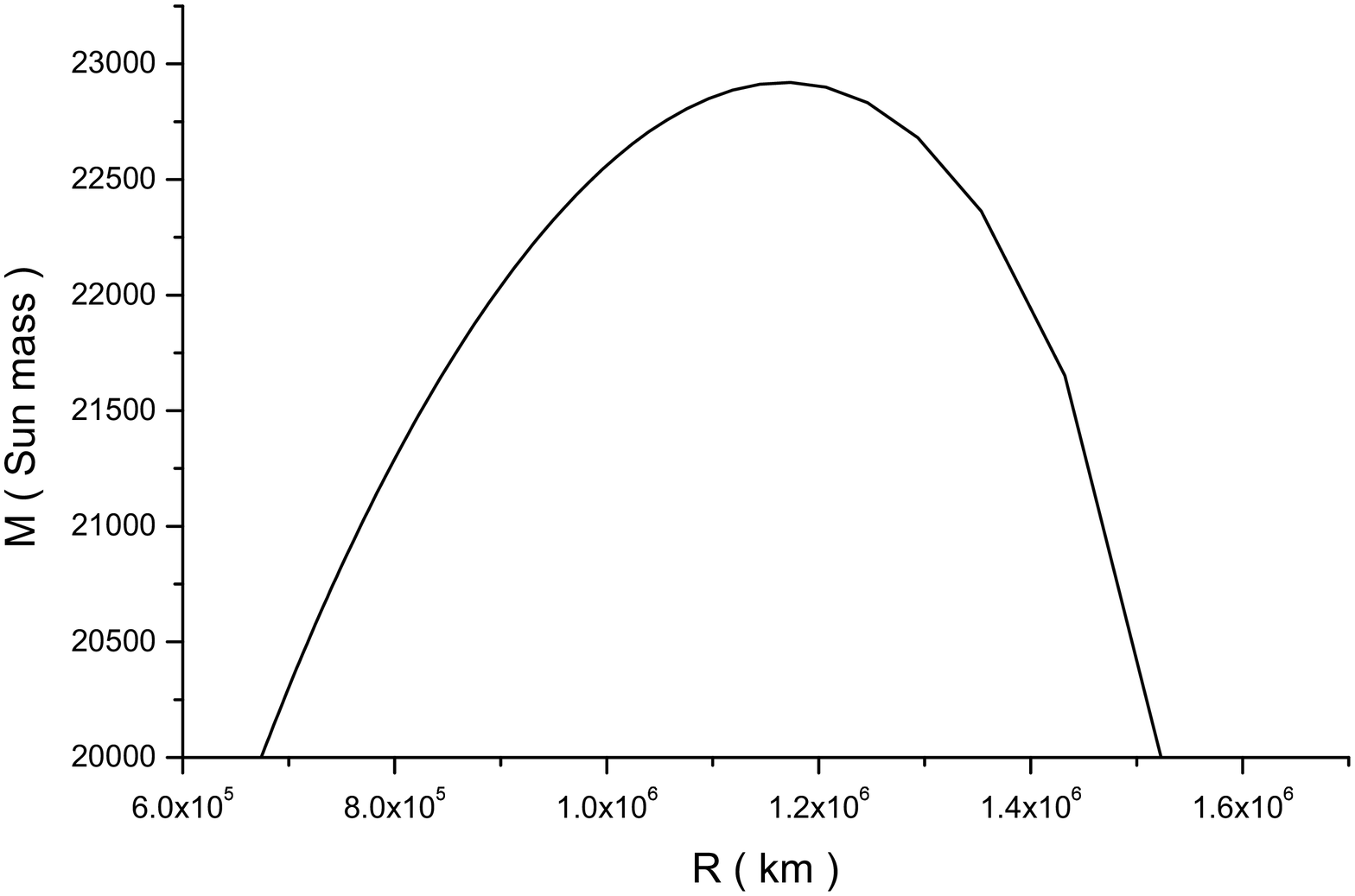}} \par}
\caption{\label{rmd}The R-M diagram for the neutrino star. }
\end{figure}

\begin{figure}
{\par\centering \resizebox*{10cm}{!}{\includegraphics{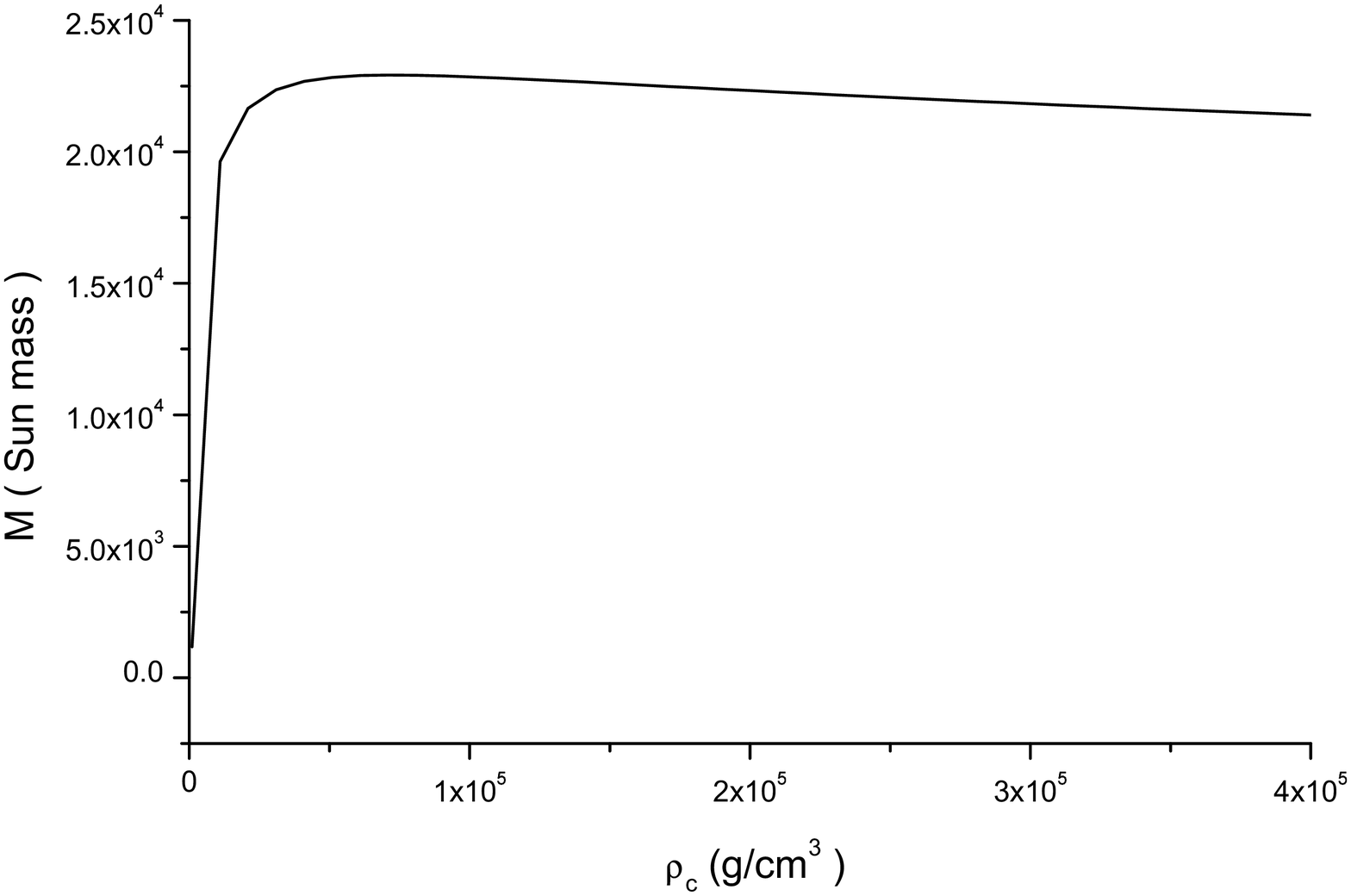}} \par}
\caption{\label{rys: 2}The neutrino star mass dependence from the 
central density \protect\protect\( \rho _{c}\protect \protect \).}
\end{figure}

When Fermi momentum reaches the second Kaluza-Klein level \( \sim m_{D}\sqrt{2} \)
the number of avaliable neutrino modes will change. For the high bulk neutrino
density in the ultrarelativistic limit all Kaluza-Klein modes should be included.
In that limit energy density of the bulk neutrinos is equal to 
\begin{eqnarray}
\chi _{bulk}(x_{F},T)=\frac{\gamma }{3\pi ^{2}}(r_{2}M_{2})\int dz\, z^{5}\{\frac{1}{(e^{(z-x_{F})/\tau }+1)}+\frac{1}{(e^{(z+x_{F})/\tau }+1)}\} &  & \\
=-\frac{40\gamma }{\pi }(r_{2}M_{2})t^{6}\int ^{\infty }_{0}\frac{dz\, z^{5}}{e^{(z-x_{F})/\tau }+1}\left\{ Li_{6}(e^{x_{F}/\tau })+Li_{6}(e^{-x_{F}/\tau })\right\} . &  &
\end{eqnarray}
In the ultrarelativistic limit the equation of state is 
\begin{equation}
\phi _{bulk}(x_{F},T)=\frac{1}{5}\epsilon _{bulk}(x_{F},T).
\end{equation}
 The equations (\ref{teq1},\ref{teq2}) are easy integrated numerically, For
example, for the neutron star with the central density \( \rho _{c}=\, 10^{4}\, g/cm^{3} \)
and temperature \( T=50\, keV \) the star density and pressure profile are
presented on the Fig.\ref{rfig1} and Fig. \ref{rfig2}. Similarly  to a structure
of white dwarf massive sterile neutrinos like ions contribute to the density
of the star , while the massles neutrinos like electrons contribute to the pressure.
This feature will be more visible with the increasing temperature. 
 The parameters of the maximum mass configuration are: 
\begin{equation}
M_{max}=2.3\times 10^{4}\, M_{\odot },\, \, \, R=1.2\times 10^{6}\, km.
\end{equation}
 This fact is easy to notice on the mass-radius diagram (Fig. \ref{rmd}). In
general, we have all family of neutrino stars, depending on growing neutrino
Fermi momentum. This result concerns the Ferni momentum below the second Kaluza-Klein
level \( \sim m_{D}\sqrt{2} \) . The ulrarelativistic limit gives higher masses
\( M_{max}\sim 1.6\times 10^{6}\, M_{\odot } \). The density profile in this
limit is presented on the Fig \ref{rfig1} (magenta). In limit of bulk neutrino
ball (Appendix I) the neutrino star properties are presented in Table \ref{tab2}. 
 and the neutrino star mass dependence from the central density \( \rho _{c} \)
(Fig. \ref{rys: 2}).

\section*{Appendix}
The Einstein equations in in the six-dimensional spacetime (the metric tensor
(\ref{a2},\ref{a5}) ) can be written as
\begin{equation}
e^{-\lambda (r)}(\frac{\nu '(r)}{r}+\frac{1}{r^{2}}-2\frac{\xi '(r)^{2}}{f_{0}^{2}})-\frac{1}{r^{2}}=\frac{8\pi G_{6}}{c^{4}}e^{-2\xi (r)/f_{0}}P(r),
\end{equation}
\begin{equation}
\label{el1}
e^{-\lambda (r)}(-\frac{\lambda '(r)}{r}+\frac{1}{r^{2}}+2\frac{\xi '(r)^{2}}{f_{0}^{2}})-\frac{1}{r^{2}}=-\frac{8\pi G_{6}}{c^{2}}e^{-2\xi (r)/f_{0}}\rho (r).
\end{equation}
The Einstein equations give also the equation for the dilaton field: 
\begin{eqnarray}
 & \xi ''(r)+\frac{2}{r}\xi '(r)-\frac{1}{2}\xi '(r)(\lambda '(r)-\nu '(r))-\frac{1}{f_{0}}\xi '(r)^{2}+ & \\
 & f_{0}(\frac{1}{2r^{2}}(e^{\lambda (r)}-1)+\frac{1}{2r}(\lambda '(r)-\nu '(r))+\frac{1}{8}(\lambda '(r)-\nu '(r))\nu '(r)- & \nonumber \\
 & \frac{1}{4}\nu ''(r)+\frac{4\pi G_{6}}{c^{4}}e^{\lambda (r)-2\xi (r)/f_{0}}\, P_{s}(r)). & \nonumber 
\end{eqnarray}
The continuity equation \( T^{MN}_{;M}=0 \) gives
\begin{equation}
\nu '(r)=-\frac{2P'(r)}{(P+c^{2}\rho )}-2\frac{\xi '(r)}{f_{0}}\, \frac{(P-2P_{s}-c^{2}\rho )}{(P+c^{2}\rho )}.
\end{equation}
We assume that the energy-momentum tensor has a diagonal form \( diag(T_{\mu \nu })=\{\varepsilon ,P,P,P,P_{s},P_{s}\} \).
Integrating of equation (\ref{el1}) yields we can write:
\begin{equation}
e^{-\lambda (r)}=\frac{\int \frac{dr}{r}((1-\frac{8\pi G}{c^{2}}r^{2}\rho (r)\, e^{-\frac{2\xi (r)}{f_{0}}})\, e^{\int dr(\frac{1}{r}+\frac{2r}{f_{0}^{2}}\xi '(r)^{2})})}{\exp (\int dr(\frac{1}{r}+\frac{2r}{f_{0}^{2}}\xi '(r)^{2}))}.
\end{equation}
 If we define 
\begin{equation}
e^{-\lambda (r)}=1-\frac{2G_{6}}{c^{2}r}m(r),
\end{equation}
 then we can obtain the generalized Oppenheimer-Tolman-Volkoff equation:
\begin{eqnarray}
\frac{dP(r)}{dr}=-\frac{G_{6}}{r^{2}}\frac{(\rho (r)+\frac{P(r)}{c^{2}})(m(r)+\frac{4\pi }{c^{2}}P(r)r^{3}\, e^{-2\xi (r)/f_{0}})}{(1-\frac{2G_{6}}{c^{2}r}m(r))}+ &  & \\
(\frac{\xi '(r)^{2}}{f_{0}^{2}}-\frac{\xi '(r)}{^{f_{0}}}\frac{(P-2P_{s}-c^{2}\rho )}{(P+c^{2}\rho )})\, (c^{2}\rho +P). &  & 
\end{eqnarray}
 In vacuum \( P=0\,, \, P_{s}=0 \), of course we have well known Schwarzschild
solution:
\begin{eqnarray*}
e^{-\lambda (r)}=(1-\frac{r_{g}}{r}), &  & \\
e^{\nu (r)}=(1-\frac{r_{g}}{r}), &  & \\
\xi (r)=0. &  & 
\end{eqnarray*}
Neglecting the dilaton field \( \xi (r) \) the Oppenheimer-Tolman-Volkoff (\ref{teq1}).

The theory of neutrinos, bound by gravity, can be easily sketched considering
a Thomas-Fermi model for fermions \cite{bilic}. We can set the Fermi energy
equal to the gravitational potential which binds the system, and see that the
number density is a function of the gravitational potential. Such a gravitational
potential will obey a Poisson equation, where neutrinos (and anti-neutrinos)
are the source term. Including gravity the local equilibrium condition demands
\begin{equation}
\frac{1}{r^{2}}\frac{d}{dr}(\frac{r^{2}}{\rho (r)}\frac{dp}{dr})=-4\pi G_{N}\rho (r).
\end{equation}
Inside the ball the pressure of the massive bulk neutrinos is \( P=\rho /5 \)
while \( P=\rho /3 \) for the brane neutrinos in the relativistic limit in
high temperature. Defining 
\begin{equation}
\rho (r)=\rho _{c}e^{\varphi }
\end{equation}
 with 
\begin{equation}
\kappa =24\pi G_{N}\rho _{c}
\end{equation}
 we have the Liuoville equation 
\begin{equation}
\triangle \varphi =-\kappa e^{\varphi }.
\end{equation}
 The Laplace operator is 
\begin{equation}
\triangle =\frac{1}{r^{2}}\frac{d}{dr}(r^{2}\frac{d}{dr})+...
\end{equation}
 Using the thin-wall approximation one can obtain the following expression 
\begin{equation}
\label{wz_1}
F_{0}(r)=e^{\phi }=\frac{1}{{\cosh }^{2}(\frac{\sqrt{2\kappa }}{2}r)}.
\end{equation}
 The equation (\ref{wz_1}) allows to estimate the mass of the neutrino ball
which is given by the dependence 
\begin{equation}
M(R)=4\pi \rho _{c}\int _{0}^{\infty }\frac{drr^{3}}{{\cosh }^{2}(\frac{\sqrt{2\kappa }}{2}r)}
\end{equation}
 with the radius scale 
\begin{equation}
r_{0}=\sqrt{\frac{\kappa }{2}}.
\end{equation}
The bulk neutrino density profile (\ref{wz_1}) is presented on the Fig. \ref{rfig1}
(the dotted line).
\begin{table}
{\centering \begin{tabular}{|c|c|c|}
\hline 
\( \rho _{c}\, (g/cm^{3}) \)&
\( R\, (km) \)&
\( M\, (M_{\odot }) \)\\
\hline 
\hline 
\( 10^{3} \)&
\( 5.9\, 10^{6} \)&
\( 1.1\, 10^{6} \)\\
\hline 
\( 10^{4} \)&
\( 1.8\, 10^{6} \)&
\( 3.5\, 10^{5} \)\\
\hline 
\( 10^{5} \)&
\( 5.9\, 10^{5} \)&
\( 1.1\, 10^{5} \)\\
\hline 
\end{tabular}\par}

\caption{\label{tab2}The bulk neutrino star properties in the neutrino ball approximation. }
\end{table}


\begin{thebibliography}{10}
\bibitem{1}N. Arkani-Hamed, S. Dimopoulos, and G. Dvali, Phys. Lett. 
\textbf{B429}, 263 (1998). 
\bibitem{4}I. Antoniadis, N. Arkani-Hamed, S. Dimopoulos, and G. Dvali, 
Phys. Lett. \textbf{B436}, 257 (1998). 
\bibitem{ADD3}N. Arkani-Hamed, S. Dimopoulos, and G. Dvali,
Phys. Rev. \textbf{D59}, 086004 (1999).
\bibitem{L1}J. Lykkin, Talk at the International Workshop on Phenomenological Aspects of
Superstring Theories (PAST97), Trieste, Italy, 2-4 October, 1997.
\bibitem{DDG1}K.R. Dienes, E. Dudas, and T. Gherghetta, Phys. Lett. 
{\bf B436}, 55 (1998).
\bibitem{hd}B.Holdom, Phys. Rev. \textbf{D36}, 1000 (1987); 
\bibitem{torr}D.F.Torres, S.Cappozziello, G. Lambiase, 
\textit{A supermassive scalar star at the Galactic Center?,} 
\url{http://xxx.lanl.gov/abs/astro-ph/0004064}.
\bibitem{nk}N. Kan, K. Shiraishi, \textit{Fermion star with an Extra Dimension},
\url{http://xxx.lanl.gov/abs/gr-qc/0001027}.
\bibitem{tap}T.Appelquist, A.Chodos, P.G.O. Freund \textit{Modern Kaluza-Klein Theories},
Addison-Wesley Publishing Comp. Meno Park, 1987. 
\bibitem{dinest}K. R. Dienes, E. Dudas, and T. Gherghetta, \emph{Neutrino oscillations without
neutrino masses or heavy mass scales: A higher-dimensional seesaw mechanism,} 
Nucl. Phys. {\bf B557} 25 (1999);
\url{http://xxx.lanl.gov/abs/hep-ph/9811428},\url{http://xxx.lanl.gov/abs/hep-ph/9811428}.
\bibitem{ADDM}N. Arkani-Hamed, S. Dimopoulos, G. Dvali, and J. March-Russell, 
\textit{Neutrino masses from large extra dimensions}, 
\url{http://xxx.lanl.gov/abs/hep-ph/9811448},
\url{http://xxx.lanl.gov/abs/hep-ph/9811448}.
\bibitem{dvali}G. Dvali and A. Y. Smirnov, 
\textit{Probing large extra dimensions with neutrinos}, 
Nucl. Phys. {\bf B563} 63 (1999),\url{http://xxx.lanl.gov/abs/hep-ph/9904211}.
\bibitem{BCS}R. Barbieri, P. Creminelli, and A. Strumia, 
\textit{Neutrino oscillations from large extra dimensions}, 
\url{http://xxx.lanl.gov/abs/hep-ph/0002199}.
\bibitem{hall}L.J. Hall, D. Smith, \textit{Constraints on Theories with Large Extra Dimensions,}
\url{http://xxx.lanl.gov/abs/hep-ph/9904267}, Phys. Rev. {\bf D60}, 085008
(1999).
\bibitem{karmen}A. Lukas, A. Romanino, \textit{A Brane-World Explanation of the KARMEN Anomaly},
\url{http://xxx.lanl.gov/abs/hep-ph/0004130}.
\bibitem{rm}R.Ma\'{n}ka, I.Bednarek, D.Karczewska, 
\textit{The neutrino ball in the standard model}. 
\url{http://xxx.lanl.gov/abs/astro-ph/9304007}; Phys. Scr. {\bf 52}, 36
(1995);
R.Ma\'{n}ka, D.Karczewska, Z. Phiz. {\bf C57}, 417 (1993). 
\bibitem{tsi}Tsiklauri D., and Viollier R. D., ApJ., 
{\bf 500}, 591 1998; 
\bibitem{fet}A.L.Fetter, J.D.Walecka, 
\textit{Quantum Theory of Many-Particle Systems}, McGraw-Hill, New York, 1971.
\bibitem{toki}K.Sumiyoshi, H.Toki, ApJ, {\bf 422}, 700 1994.
\bibitem{bilic}N.Bilic, D.Tsiklauri, R.Viollier, 
Prog. Part. Nucl. Phys. {\bf 40}, 17 (1998).
\end{thebibliography}
\end{document}